\documentclass[twoside]{article}
\usepackage{fleqn,espcrc2}
\usepackage{amsmath}

\usepackage[dvips]{graphicx}

\title{Probing the QCD vacuum with overlap fermions
\thanks{Preprint FSU-CSIT-01-61, JLAB-THY-01-37. Talk given by U.M.~Heller at
the Workshop on Lattice Hadron Physics ``LHP2001'', July 9 -- 18, 2001, Cairns,
Australia.}}
\author{Robert G. Edwards
\address{Jefferson Lab, 12000 Jefferson Avenue, MS 12H2,
Newport News, VA 23606, USA}
and Urs M. Heller
\address{CSIT, Florida State University, Tallahassee, FL 32306-4120, USA}
} 

\begin{document}

\begin{abstract} 
We use low lying eigenvectors of the overlap-Dirac operator as a probe
of the QCD vacuum. If instantons play a significant role one would expect
the low lying eigenmodes of the overlap-Dirac operator to consist mainly
of the mixed ``would be zero modes''. Then, the eigenmodes should exhibit
local chirality. Studying a recently introduced local chirality
parameter, we find evidence supporting this picture.
\end{abstract}

\maketitle

\section{Introduction}

Understanding the QCD vacuum has been a goal of lattice QCD for many
years. This understanding, in turn, should illuminate the
mechanisms of confinement and chiral symmetry breaking. A good candidate
for the source of chiral symmetry breaking are topological fluctuations
of the gauge fields associated with instantons and anti-instantons.
Through mixing, the ``would be zero modes'' from the instantons and
anti-instantons become the small modes that lead, when present with a
finite density, to chiral symmetry breaking, since $\langle \bar\psi
\psi \rangle = \pi \rho(0^+)$, with $\rho(0^+)$ the density of the small
modes. Many of the low energy properties of QCD can be explained
phenomenologically by the interactions of the fermions with instantons
and anti-instantons \cite{inst_liquid,diak}. To our knowledge,
confinement, though, is not one of those properties.

There is a problem, though, with investigating the topological content of
Monte Carlo generated gauge configurations. Ultraviolet fluctuations
(noise) dominate these configurations. Pure gauge probes of the vacuum
therefore need filters, like smearing or cooling, to get rid of the
UV noise and make it possible to uncover the underlying topological
fluctuations \cite{Teper_Pisa}. But do such filters change the physics?
Fermions, on the other hand, should provide a more physical filter,
provided lattice artifacts do not impede the chiral and topological
properties. Since a few years such fermions are known, the overlap
fermions of Neuberger \cite{Herbert}.

Global topology, {\it i.e.} the topological susceptibility $\chi_t$ has
already been successfully computed from the number and chirality of
the zero modes. This was actually done, completely equivalently, from
studying the spectral flow of the underlying Wilson Dirac operator
\cite{Q_flow}.

Here we want to address local properties. If instanton and anti-instanton
like fluctuations are important, then the low-lying non-zero eigenmodes
should have significant contributions from mixed ``would-be zero modes''.
Locally, in their peaks, these low eigenmodes should therefore be chiral,
since zero modes are chiral, even though the integrated chirality of
all non-zero modes is exactly zero. This can be measured with the
``local chirality'' parameter, $X(x)$, defined by \cite{HIMT}
\begin{equation}
\tan \Bigl( \frac{\pi}{4} (1+X(x)) \Bigr) = \biggl(
 \frac{\psi^\dagger_L(x) \psi_L(x)} {\psi^\dagger_R(x) \psi_R(x)}
 \biggr)^{1/2} ~,
\label{eq:chir_1}
\end{equation}
where $\psi_L(x)$ and $\psi_R(x)$ are the left- and right-handed components
of the eigenmode.

It should be noted, however, that Witten pointed out an inconsistency
between instanton based phenomenology and large-$N_c$ QCD \cite{Witten}.
Instantons would produce an $\eta^\prime$ mass that vanishes exponentially
for large $N_c$, while considerations based on large-$N_c$ chiral dynamics
suggest that the $\eta^\prime$ mass should be of order $1/N_c$. The
topological charge fluctuations then should be associated not to instantons
but rather some other, confinement-related vacuum fluctuations. And the
strong attraction produced by these confinement-related fluctuations
would also induce the breaking of chiral symmetry. Then there is no reason
that the low-lying modes should be chiral in their peaks. Using non-chiral
Wilson fermions at a lattice spacing of $a \simeq 0.17$ fm Horv\'ath
{\em et al.}~\cite{HIMT} found that the local chirality $X(x)$ was not
peaked near $\pm 1$. We use chiral overlap fermions instead to avoid the
potentially large ${\cal O}(a)$ lattice artifacts from the explicit
chiral symmetry breaking term of Wilson fermions.

\section{Setup}

For our study we consider the standard massless overlap-Dirac operator
\cite{Herbert}
\begin{equation}
D(0) = \frac{1}{2} \bigl[ 1 + \gamma_5 \epsilon(H_w(M)) \bigr]
\label{eq:Dov}
\end{equation}
with $H_w(M) = \gamma D_w(-M)$ and $D_w(M)$ the usual Wilson-Dirac operator.
Then, $H^2(0) = D^\dagger(0) D(0)$ commutes with $\gamma_5$ and can
be simultaneously diagonalized~\cite{EHN_practical,ParalComp}. $H^2(0)$
can have zero eigenvalues with chiral eigenmodes, which are also
eigenmodes of $D(0)$, due to global topology. The non-zero eigenmodes of
$H^2(0)$ are doubly degenerate and have opposite chirality,
\begin{equation}
H^2(0) \psi_{\uparrow,\downarrow} = \lambda^2 \psi_{\uparrow,\downarrow}
 ~,\qquad \gamma_5 \psi_\uparrow = \psi_\uparrow ~,
 \gamma_5 \psi_\downarrow = - \psi_\downarrow ~,
\end{equation}
with $0 < \lambda \le 1$.
The non-zero (right) eigenmodes of $D(0)$ are then easily obtained as
\begin{equation}
\psi_\pm = \frac{1}{\sqrt{2}} \Bigl( \psi_\uparrow
 \pm i \psi_\downarrow \Bigr)
\end{equation}
with eigenvalues
\begin{equation}
\lambda_\pm = \lambda^2 \pm i \lambda \sqrt{1 - \lambda^2} ~.
\end{equation}
The ``chirality'' parameter of eq.~(\ref{eq:chir_1}) is then given by
\begin{equation}
\tan \Bigl( \frac{\pi}{4} (1+X(x)) \Bigr) = \biggl(
 \frac{\psi^\dagger_\downarrow(x) \psi_\downarrow(x)}
 {\psi^\dagger_\uparrow(x) \psi_\uparrow(x)} \biggr)^{1/2} ~,
\label{eq:chir_2}
\end{equation}
and is equal for the two related modes $\psi_\pm$. In the following they
will be therefore counted as one mode.
The exact zero modes, of course, have $X(x) \equiv \pm 1$.

We computed low-lying eigenmodes $\psi_{\uparrow,\downarrow}$ of $H^2(0)$
with the Ritz functional algorithm of Ref.~\cite{Ritz}. For the sign
function $\epsilon(H_W)$ in eq.~(\ref{eq:Dov}) we used the optimal
rational approximation of \cite{EHN_practical,ParalComp} with ``projection
of low-lying eigenvectors of $H_w$'' to ensure sufficient accuracy.

\section{Results}

For a more extensive and detailed discussion of our results see
\cite{EH_fluc}. Here, we concentrate on the results obtained on gauge
field configurations generated with the Iwasaki action~\cite{IW}.
At comparable lattice spacing, the implementation of overlap fermions
on such configurations is about a factor 3 faster than on Wilson
action configurations~\cite{EH_fluc}. We computed the 20 eigenmodes
$\psi_{\uparrow,\downarrow}$ of $H^2(0)$ with smallest $\lambda$, with
the Wilson-Dirac mass in the overlap-Dirac operator set to $M=1.65$.
Six ensembles were considered corresponding to three different lattice
spacings and three different volumes, in physical units
(see Table \ref{tab:Q_Qsq}).

\begin{table}
\caption{The gauge ensembles. Listed are the volume in lattice units and
in units of fm$^4$, the number of configurations analyzed, the average
number of zero modes and the average of the square of the global topological
charge as obtained from the number of zero modes.}
\label{tab:Q_Qsq}
\vspace{2mm}
\begin{tabular}{|l|r|r|l|l|l|}
\hline
 $\beta$ & $V$ & $V$ & $N$ & $\langle |Q| \rangle$ & $\langle Q^2 \rangle$  \\
 & [$a^4$] & [fm$^4$] & & & \\
\hline
 2.2872 &  $6^3 \times 12$ &  2.2 & 100 & 1.0(1) & 1.7(2)  \\
 2.2872 &  $8^3 \times 16$ &  7.0 &  69 & 2.0(2) & 5(1)    \\
 2.2872 & $12^3 \times 16$ & 23.6 &  10 & 4.2(9) & 25(8)   \\
 2.45   &  $8^3 \times 16$ &  2.0 & 197 & 1.0(1) & 1.8(2)  \\
 2.45   & $12^3 \times 16$ &  6.7 &  47 & 2.3(4) & 8(1)    \\
 2.65   & $12^3 \times 16$ &  2.1 &  50 & 1.3(2) & 2.4(6)  \\
\hline
\end{tabular}
\vspace*{-2mm}
\end{table}

\begin{figure}
\vspace{2.7cm}
\centerline{\includegraphics[width=3.0in]{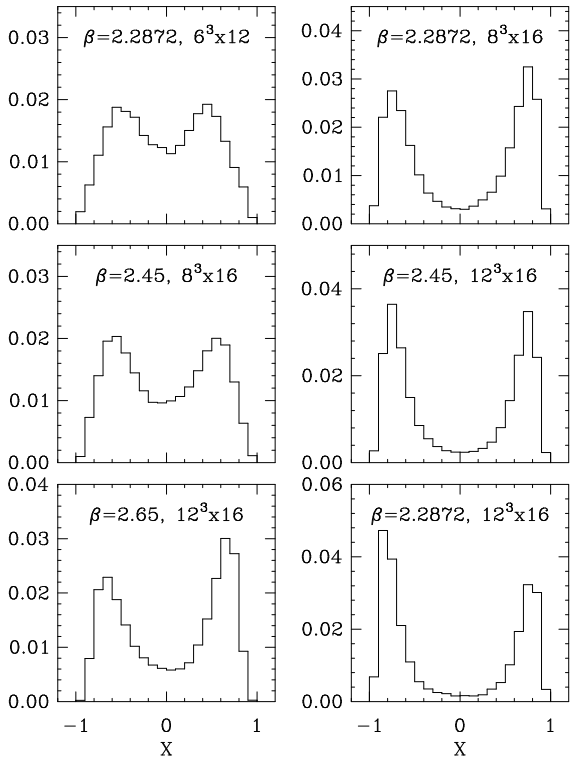}}
\vspace*{-5mm}
\caption{Chirality histograms for the lowest two non-zero modes of the
overlap-Dirac operator at the 2.5\% sites with the largest
$\psi^\dagger \psi(x)$.}
\label{fig:hist_IW_2l_2p5pc}
\vspace*{-2mm}
\end{figure}

The chirality histograms of the two lowest non-zero modes at the 2.5\%
of sites with largest $\psi^\dagger \psi(x)$ for all six ensembles
are shown in Fig.~\ref{fig:hist_IW_2l_2p5pc}. The physical
volume of the systems shown in the left column are approximately the
same, 2 fm$^4$. We see that the lowest non-zero modes tend to become more
chiral in their peaks as the lattice spacing is decreased. This strongly
suggests that the behavior observed here will survive the continuum limit.
Comparing the three histograms for $\beta=2.2872$ having equal lattice
spacing, we see a dramatic dependence on the physical volume.
In an instanton liquid picture of the vacuum, the number of instantons and
anti-instantons, and hence the number of almost zero modes grows linearly
with the volume. Then it is not surprising that the lowest modes become
increasingly chiral in their peaks with increasing volume.

\begin{figure}
\vspace*{-5mm}
\centerline{\includegraphics[width=3.0in,height=1.5in]{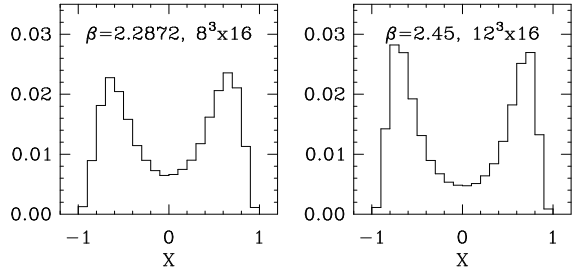}}
\caption{Chirality histogram for the lowest six non-zero modes at
the 2.5\% sites with the largest $\psi^\dagger \psi(x)$ on the two lattices
with volume of about 7 fm$^4$.}
\label{fig:hist_IW_6l_2p5pc}
\end{figure}

\begin{figure}
\vspace*{-5mm}
\centerline{\includegraphics[width=1.4in]{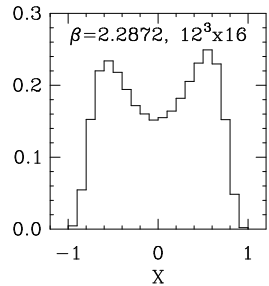}}
\vspace*{-5mm}
\caption{Chirality histogram for all the non-zero modes out of the 20
lowest modes at the 20\% sites with the largest $\psi^\dagger \psi(x)$
on the lattice at $\beta=2.2872$ with the largest volume.}
\label{fig:hist_229LL_all_2p5pc}
\vspace*{-2mm}
\end{figure}

In Fig.~\ref{fig:hist_IW_6l_2p5pc} we show the chirality histogram of
the lowest six non-zero modes on the two systems with physical volume
of about 7 fm$^4$ and in Fig.~\ref{fig:hist_229LL_all_2p5pc} the chirality
histogram from all non-zero modes out of the 20 lowest modes that we
computed (there are between 11 and 20 non-zero modes per configuration)
for the system with the largest physical volume. Here we kept the 20\%
sites with largest $\psi^\dagger \psi(x)$.
Comparing with the histograms in the left column of 
Fig.~\ref{fig:hist_IW_2l_2p5pc} indicates that the number of non-zero
modes with similar chirality histograms grows roughly like the physical
volume so that there is a finite density of modes which are chiral in their
peaks. This is good evidence that we are not observing a finite volume
effect that would disappear in the infinite volume limit.

\begin{table}
\caption{The average of the sum of $\psi^\dagger \psi(x)$ over the 2.5\%
of sites kept for the chirality histograms as  a percentage of the sum over
all sites for some low-lying eigenvectors for the various gauge field 
ensembles.}
\label{tab:psi_sq}
\vspace{2mm}
\begin{tabular}{|l|r|l|l|l|l|l|}
\hline
 $\beta$ & $V$ [$a^4$] & ev 1 & ev 2 & ev 20 \\
\hline
 2.2872 &  $6^3 \times 12$ &  8.9(2) &  7.8(2) &  5.1(1) \\
 2.2872 &  $8^3 \times 16$ & 11.5(3) & 10.6(2) &  6.2(2) \\
 2.2872 & $12^3 \times 16$ & 12.9(5) & 13.4(6) & 11.3(5) \\
 2.45   &  $8^3 \times 16$ &  9.9(2) &  8.7(1) &  6.0(1) \\
 2.45   & $12^3 \times 16$ & 12.4(3) & 11.5(3) &  7.8(3) \\
 2.65   & $12^3 \times 16$ & 10.4(3) &  9.2(2) &  6.9(3) \\
\hline
\end{tabular}
\end{table}

In the chirality histograms shown we typically considered the contribution
from the 2.5\% sites with the largest $\psi^\dagger \psi(x)$. In
Table~\ref{tab:psi_sq} we list the sum of $\psi^\dagger \psi(x)$ over
those sites as a percentage of the sum over all sites for some low-lying
eigenvectors for all the gauge field ensembles considered. The more chiral
the eigenvectors are on those 2.5\% sites, the higher is the percentage of the
eigenvector contained on those sites.

\section{Abelian gauge theories}

\begin{figure}
\vspace*{-5mm}
\centerline{\includegraphics[width=1.4in]{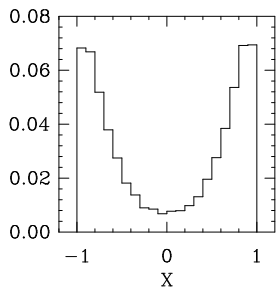}}
\vspace*{-5mm}
\caption{Chirality histogram for the lowest ten non-zero modes of the 
overlap-Dirac operator at the 6\% sites with the largest 
$\psi^\dagger \psi(x)$ on $24^2$ U(1) configurations at $\beta=1.9894$.}
\label{fig:hist_u1d2_10l_6pc}
\vspace*{-2mm}
\end{figure}

To contrast the results for quenched QCD of the previous section we
considered the chirality parameter also for quenched U(1) theories in
two and four dimensions. In 2d
topology plays a crucial role and one expects local chirality in the peaks
of low-lying modes. This is clearly seen in Fig.~\ref{fig:hist_u1d2_10l_6pc},
where the 10 lowest non-zero modes have been kept.

\begin{figure}
\vspace*{-5mm}
\centerline{\includegraphics[width=3.0in,height=1.5in]{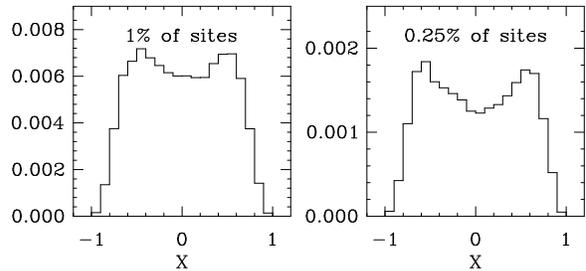}}
\caption{Chirality histogram for the lowest two non-zero modes of the
overlap-Dirac operator at the 1\% (left panel) and 0.25\% (right panel)
sites with the largest $\psi^\dagger \psi(x)$ on $8^4$ U(1) configurations
at $\beta=0.9$.}
\label{fig:hist_u1d4_2l_comp}
\vspace*{-1mm}
\end{figure}

In 4d U(1), in the confined phase, chiral symmetry is spontaneously broken,
and a finite density of near zero modes exists. But there are no instantons,
and so one would not expect the near zero modes to be dramatically chiral even
in their peak regions. There do exist exact zero modes of the overlap-Dirac
operator, which again, of course, are chiral, although their origin is not
completely understood~\cite{U1_ov}. We analyzed, stored eigenmodes from
Ref.~\cite{U1_ov}. The chirality histogram for the lowest two non-zero modes
is shown in Fig.~\ref{fig:hist_u1d4_2l_comp}. We note that while there is
some mild indication of chirality peaking, the proportion of sites showing
this behavior is much reduced compared to the SU(3) case and appears to be
of a qualitatively different nature than seen before. 
Also, the sum of $\psi^\dagger \psi(x)$ over the sites kept for the
histograms of Fig.~\ref{fig:hist_u1d4_2l_comp} where almost identical
for all 12 lowest eigenmodes (4.0\% and 1.20\%, respectively; for the
2.5\% of sites with largest $\psi^\dagger \psi(x)$ the sum is about 8.4\%
for all eigenmodes as compared to the non-abelian case in
Table~\ref{tab:psi_sq}).
In the U(1) case the (near) lack of peaking could conceivably come about
from the scenario outlined by Horv\'ath, {\em et. al.} -- namely from the
confinement inducing vacuum fluctuations.

\section{Conclusions}

By studying the local chirality parameter, introduced in Ref.~\cite{HIMT},
for low-lying non-zero eigenmodes of the overlap-Dirac operator, we
investigated the topological content of the QCD vacuum.
Considering several volumes and lattice spacings, we found convincing
evidence for chirality of the low lying modes in their peak region.
Our results give evidence that the number of modes which are chiral in
their peaks grows linearly with the volume so that there is a {\it finite
density} of such modes, and that the chirality of the modes becomes more
pronounced as the lattice spacing is decreased. Therefore, our observations
should remain valid in the continuum limit. Our observations are
consistent with the topological fluctuations being instanton dominated
and the low-lying eigenmodes being the mixed ``would be zero modes''
due to the instantons and anti-instantons.

Our findings confirm the results of Ref.~\cite{DH}, where a different
version of overlap fermions coupled to smoothened gauge fields were used,
and extend them by a careful study of finite lattice spacing and volume
effects. Our results, on the other hand, contradict the conclusions of
Horv\'ath {\em et al.}~\cite{HIMT} who studied the chirality parameter
with Wilson fermions at a large lattice spacing.
That the results of Ref.~\cite{HIMT} are strongly affected by lattice
artifacts was confirmed in a recent paper of
Hip {\it et al.}~\cite{WC_imp}. These authors considered an improved
chirality parameter for Wilson fermions, clover improved Wilson fermions,
and lattices at smaller lattice spacing than Ref.~\cite{HIMT}. With
these reductions of lattice artifacts, they conclude that instanton
dominance of topological charge fluctuations is not ruled out by the
response of (improved) Wilson fermions. Together with the results from
overlap fermions presented here and the lack of significant
chirality enhancement in the 4-d U(1) model where instantons should not
exist, the case for instanton domination in 4-d SU(3) gauge theory, 
as measured by the chirality parameter, becomes even more
compelling. However, for an opposite point of view see Ref.~\cite{Horv_ov}.

\section*{Acknowledgements}

We thank the organizers for an exciting workshop.
RGE was supported by DOE contract DE-AC05-84ER40150 under which the
Southeastern Universities Research Association (SURA) operates the
Thomas Jefferson National Accelerator Facility (TJNAF).  UMH was
supported in part by DOE contract DE-FG02-97ER41022. The computations
were performed on the JLab/CSIT QCDSP at JLab, 
the JLab/MIT workstation cluster and the CSIT workstation cluster.

\end{document}